\documentclass[notitlepage]{revtex4-1}

\usepackage{lipsum} 
\usepackage{amsfonts, amsmath, amsthm, amssymb} 
\usepackage{graphicx} 
\usepackage{booktabs} 
\usepackage{wrapfig} 
\usepackage[labelfont=bf]{caption} 
\usepackage[top=0.7in,bottom=0.7in,left=0.4in,right=0.4in]{geometry} 
\usepackage{lastpage}
\usepackage{xcolor}

\begin{document}

\title{\large microRNA-mediated noise processing in cells: a fight or a game?}

\author{Elsi Ferro$^\ast$}
\affiliation{Italian Institute for Genomic Medicine, Italy}
\author{Chiara Enrico Bena$^\ast$}
\affiliation{Italian Institute for Genomic Medicine, Italy}
\author{Silvia Grigolon$^\dagger$}
\affiliation{The Francis Crick Institute, 1, Midland Road, London NW1 1AT, UK}
\author{Carla Bosia$^\dagger$}
\affiliation{Italian Institute for Genomic Medicine, Italy}
\affiliation{Department of Applied Science and Technology, Politecnico di Torino, Italy}
\thanks{$^\ast$ These authors contributed equally to this work}
\thanks{$^\dagger$ Corresponding authors}

\maketitle
\vspace{-0.8cm}
\begin{center}
\end{center}

\section*{Abstract}
In the past decades microRNAs (miRNA) have much attracted the attention of researchers at the interface between life and theoretical sciences for their involvement in post-transcriptional regulation and related diseases. Thanks to the always more sophisticated experimental techniques, the role of miRNAs as ``noise processing units'' has been further elucidated and two main ways of miRNA noise-control have emerged by combinations of theoretical and experimental studies. While on one side miRNA were thought to buffer gene expression noise, it has recently been suggested that miRNA could also increase the cell-to-cell variability of their targets. In this Mini Review, we focus on the miRNA role in noise processing and on the inference of the parameters defined by the related theoretical modelling.

\section{Introduction}

To carry out all vital functions, cells must express proteins with a high precision in time and protein numbers. Protein production, i.e. gene expression, results from the complex interactions among a high variety of molecules, among which transcription factors, genes, short and long RNAs, and ribosomes. Due to the inherent stochasticity of chemical reactions, gene expression is naturally highly noisy, thus leading to a wide range of possible values of produced proteins. Contrary to expectations, Poissonian distributions are not the standard experimental outcome for most genes and larger fluctuations in the number of transcripts are instead observed.

Amongst others, stem cells show clear examples of this.
Analysis of the expression variability landscape in pluripotent stem cells (PSCs) shows indeed that several gene transcripts display lognormal or bimodal distributions across the population \cite {Kumar2014}. In spite of the apparent uniformity of PSCs, a cell population can even contain rare subpopulations expressing markers of different cell lineages. Analysis of gene expression data collected upon perturbation of single PSCs allowed the identification of the main variability axes: the key genes that generate this heterogeneity resulted to be the main pluripotency transcription factors (PFs) that are considered to play a primary role in maintaining the pluripotency state of a cell. Indeed, the key PFs were observed to fluctuate in a reciprocally correlated manner throughout the population, and the regulatory relationships amongst them were shown to adopt different configurations depending on the cell state. Similarly, mouse embryonic stem cells (mESCs) appear to be heterogenous in their gene expression profile as well \cite {Klein2015}. This heterogeneity may indicate either reversible fluctuations or already ongoing differentiation processes. In fact, during differentiation, gene-expression correlations displayed important changes due to PFs switching off in an alternate way, thus allowing the appearance of novel cell states. Gene-expression variability might therefore play an essential role in fundamental biological processes such as cell fate decision. 

Large efforts in the past few years have been dedicated to identify the mechanisms that generate these fluctuations. Stochasticity comes as an inherent feature of ``small-number'' probabilistic phenomena, thus the interactions between small amounts of molecules, such as the reactions underlying gene expression, are intrinsically noisy. Besides, mechanisms of large-fluctuation generation could also be attributed to large variations in the state of gene-specific promoters, acting in a switch mode \cite{shahrezaei2008}. When in the on-state, the promoters lead to bursts in gene expression \cite{battich2015}, consequently increasing the variance of the final protein product. Nevertheless, gene-expression noise can also arise from factors external to the gene, that indirectly affect its function. Cell-to-cell variability such as fluctuations in the environment (e.g., thermal fluctuations), ribosome abundance and ATP availability are further sources of noise. Given this background, noise in gene expression is normally classified into \emph{intrinsic noise}, due to the inherent stochasticity of transcription, translation and decay processes, and \emph{extrinsic noise}, due to any ``external'' fluctuation that indirectly leads to expression variations \cite{swain2002intrinsic}.

A natural quantitative measure of gene-expression noise is the size of protein fluctuations compared to their mean amount \cite{swain2002intrinsic}, thus if $P(t)$ is the protein concentration at time $t$, the noise $\eta(t)$ is given by: $\eta^2(t)=\frac{\langle P(t)^2 \rangle - \langle P(t) \rangle ^2}{\langle P(t) \rangle ^2}$ (1), that is the ratio of variance to mean of the number of protein molecules per cell. By considering the expression variability of a particular gene across a cell population, Swain and colleagues suggested how noise can be mathematically decomposed into intrinsic and extrinsic contributions \cite{swain2002intrinsic}. They showed that the total noise is given by the sum in quadrature of intrinsic and extrinsic components, that is $\eta_{tot}=\sqrt{\eta_{int}^2+\eta_{ext}^2}$ (2). Different works have proposed analytical expressions for $\eta_{int}$ and $\eta_{ext}$ that depend on the sources of fluctuations and on the available measurable quantities \cite{swain2002intrinsic, thomas2019intrinsic, singh2013quantifying}.

On the experimental side, the issue of how intrinsic and extrinsic noise contributions can be discriminated has been initially addressed by Elowitz and co-workers \cite{elowitz2002stochastic}. By considering two identical gene copies present in the same cell, they measured their protein products simultaneously. Because the gene copies are both exposed to the same intracellular environment, one can assume that their variability is solely due to intrinsic factors. Thus by tagging the two genes with distinguishable fluorescent probes and measuring the average deviation between the two protein amouts over a cell population, intrinsic noise can be quantified and extrinsic noise will then follow from (2) \cite{swain2002intrinsic}. To date, this simple ``dual-reporter'' framework has been pursued by a number of studies aimed at measuring noise in gene expression both \textit{in vivo} \cite{Schmiedel2015, baudrimont2019contribution} and within \textit{in silico} simulations \cite{vazquez2017extrinsic}.
Over the identification of gene expression noise sources, increasing research has committed to the understanding of how cells process these fluctuations to achieve the high precision required for life maintenance.

Recently, compartmentalisation as phase separation in cells \cite{stoeger2016} has been found to be able to play a role in the decrease of noise at the level of proteins \cite{oltsch2019}: the formation of protein aggregates increases the local number of interacting molecules, therefore decreasing the noise level.
Yet, most importantly, variability has been shown to be buffered by cells thanks to elementary gene regulatory pathways where molecules play together in tuning gene expression, i.e. network motifs. Therein, the signs of the regulatory interactions shape the target gene's response in a way that decreases the variance of the final protein outcome \cite{kontogeorgaki2017, osella2011, battich2015}. Through a number of stages, these molecular interactions are able to transform noisy signals into precise outputs \cite{laurenti2018molecular}. For example, a network where a TF both enhances the expression of a gene and simultaneously activates a repressor of the same gene, i.e. a type of feed-forward loop, has been shown to act as a noise bufferer \cite{osella2011}. In these loops, short non-coding RNAs called microRNAs (miRNAs) have often been found to mediate the repressive path by targeting transcript mRNA and preventing its translation \cite{osella2011, kim2013}. In fact, miRNAs have been found to be involved in several types of regulatory pathways \cite{ferro2019endogenous} and their functions appear to be tightly related to noise processing \cite{Herranz2010}.

All these noise-managing mechanisms are the result of the evolution process, during which cells have been selected according to an unknown fitness landscape. It is commonly believed that the minima of this landscape correspond to biological structures aimed at decreasing the noise level in gene expression \cite{lehner2008, silander2012}, in order to increase individuals' robustness against fluctuations. Yet, some studies pointed at the opposite possibility \cite{zhang2009}, i.e., to the positive selection of highly noisy genes for variability advantage. A recent study based on a combination of theory and experiments \cite{wolf2015} showed that selective pressure might even increase expression noise and the positively selected genes with elevated noise are also those highly regulated by transcription factors. On the same idea that cells do not necessarily buffer noise, a recent work showed that introduction of extrinsic noise in microRNA-mediated regulatory networks, i.e., increased variability in gene expression, can instead favour cell differentiation \cite{Bosia2017, DelGiudice2018role, DelGiudice2018stochastic, chakraborty2019}. These recent studies arouse the possibility that cells do not only buffer noise but they rather ``take advantage of stochasticity''  to optimise specific needs, e.g., cell-to-cell variability, protein number precision, information flow \cite{mancini2013, martirosyan2016}, etc. Therefore, the initial question about what mechanisms lead to noise buffering in cells should be instead changed into: what are the mechanisms that allow cells to optimise their interplay with noise? In this mini-review, we focus on the role of miRNAs in processing gene-expression variability. In the following section we will discuss how, although commonly supposed to act against noise, miRNAs are instead involved in this optimisation game. The second section is devoted to review works on the inference of miRNA-target interactions, an essential tool along the way of understanding how miRNAs deal with gene expression noise when combining theory and experiments.

\section{The role of miRNAs in noise processing}

MiRNAs are short ($\approx$ 22nt long) non-coding RNAs that work as post-transcriptional regulators by establishing and maintaining gene expression patterns \cite{cai2009, Huntzinger2011, Hausser2013}. They are encoded in nearly $1\%$ of the genome of nematodes, flies and mammals \cite{cai2009}, and they are implicated in the regulation of a variety of processes, such as the timing of developmental events, cell differentiation, proliferation and apoptosis \cite{Hwang2006}, as well as tumorigenesis and host-pathogen interactions \cite{Shahid2018}.

To carry out their regulatory roles, miRNAs bind to their mRNA targets through base pairing, with the degree of the pairing complementarity determining whether the target will undergo translational repression or mRNA increased degradation. The pairing occurs thanks to miRNA loading into RISCs, complexes involving Ago proteins that guide miRNAs to cognate mRNAs. MiRNA-dependent regulation is combinatorial, where a typical miRNA has many targets and every target is regulated by many miRNAs \cite{Hashimoto2013}. Although overrepresented in gene regulatory networks \cite{Schmiedel2017, Herranz2010, Kittelmann2019}, they exert a mild repressive role on most of their targets, with a typical fold repression smaller than two \cite{HausserZavolan2014}.
Thanks to the always more sophisticated experimental techniques, i.e. microfluidic devices, deep RNA-sequencing and single-cell transcriptome data, the role of miRNAs as noise processor units has been further elucidated and diverse ways of microRNA-mediated noise control have emerged by combinations of theoretical and experimental studies.

The first and older idea sees miRNAs playing a pivotal role in gene regulatory networks by reducing fluctuations in protein expression, thus conferring stability to the gene-expression network \cite{Herranz2010, Hilgers2010, Ebert2012, Hu2018}. Indeed gene expression may gain precision, thereby stabilising the identity of individual cells, through miRNA-mediated noise filtering \cite{Li2017}. The way miRNAs act as noise buffers is through network motifs \cite{Bokes2018}. As mentioned, interaction networks where transcription factors control the expression of the miRNAs as well as their targets (miRNA-mediated feed forward loops) are efficient in maintaining a desired expression level besides changes in gene dosage or fluctuations at the level of the master transcription factor. One of these examples is the Incoherent FeedForward Loop (IFFL), a type of circuit where a regulator TF both directly favours and indirectly inhibits the expression of a target gene through activation of a miRNA \cite{osella2011}. Theoretical modelling revealed as a useful tool for formulating predictions on the IFFL's behaviour \cite{osella2011, bosia2012, riba2014}, as these were soon experimentally confirmed \cite{kim2013, Strovas2014}.

Also regulatory modules where a miRNA and a TF mutually inhibit one another, i.e. toggle switches, have been shown to be capable of maintaining stable gene expression \cite{Herranz2010}. Siciliano and co-workers verified this ability by building a synthetic miRNA-mediated toggle-switch \cite{Siciliano2013}. They showed that such circuit is able to generate two different protein states, with the miRNA controlling the switch: in the absence of miRNAs, the cell randomly switches from one state to the other. 
However, noise appears to be endogenously controlled not only in a static way. Transcription of regulators often occurs in a fashion that alternates bursts of mRNA production and silent intervals, rather than by a constant rate of transcript accumulation \cite{li2018frequency}. A more recent investigation in the performance of regulatory elements \cite{Grigolon2016} shows how static control of protein noise is not stable. In fact, transcriptional bursting appears to be an ingredient that hampers noise reduction in feed-forward loops. An instance of this is given by the lin-4 miRNA involved in an iFFL: its pulsatile transcription allows to isolate an important developmental factor from upstream fluctuations \cite{kim2013}.

Although noise reduction seemed a hallmark of miRNA action, recent developments suggest that miRNAs may have a different effect on protein expression noise depending on the protein expression level \cite{Semrau2015}. One of the first evidences in this direction is provided by the work of Schmiedel and coworkers \cite{Schmiedel2015}, who investigated the role of miRNAs in gene-expression noise by combining mathematical modelling and single-cell reporter assays. The authors created a bidirectional plasmid reporter, depicted in Figure 1a, encoding two fluorescent versions of the same protein, ZsGreen and mCherry. The first protein is unregulated, thus its amount represents a proxy for transcriptional activity, whereas mCherry is equipped with miRNA binding sites in its 3'UTR. Since the two proteins are transcribed together, this system allows a quantitative comparison between miRNA-regulated and unregulated gene products. In order to test the effect of endogenous miRNAs, this dual reporter was transfected in mESCs and single-cell fluorescence was measured upon providing mCherry with one or multiple miR-20 binding sites. Expression fluctuations in the unregulated and regulated cases were compared for similar transcriptional activity, by binning cells according to their reporter expression level. Their results suggested that miRNA-mediated effects on noise depend on protein expression intensity: in cells with low reporter expression, mCherry noise was reduced with respect to the unregulated case, whereas in cells with high expression noise was increased. Moreover, the steepness of transition between the two regimes increased with the target complementarity and the number of miRNA binding sites. 

A theoretical model describing transcription, translation and miRNA-mediated regulation was compared to experimental data. With the noise expression decomposed as suggested by Swain et al. \cite{swain2002intrinsic}, the model predicted different effects on intrinsic and extrinsic components upon miRNA regulation: intrinsic noise is reduced with respect to the unregulated case, and the reduction depends on miRNA-mediated fold repression $r$, that is $\frac{\eta_{int}^{unreg}}{\eta_{int}^{reg}} \approx \sqrt{r}$. As suggested by Ebert et al. \cite{Ebert2012}, this reduction stems from the increased protein decay due to miRNA regulation and thus the transcription speed up required to achieve the same expression level of the unregulated case. The $\eta_{int}$ reduction was confirmed experimentally by measuring the products of two identical gene copies, one unregulated and the other equipped with miRNA binding sites. The results suggest that the reduction of intrinsic noise is an inherent feature of miRNA-mediated regulation and of post-transcriptional regulation. Bearing in mind that the overall noise increases at high expression levels, because $\eta_{int}$ is reduced, $\eta_{ext}$ must undergo an increase upon miRNA-mediated regulation. Extrinsic noise was modelled as $\eta_{ext} = \tilde{\eta}_{\mu} \times \phi$, where $\tilde{\eta}_{\mu}$ is the miRNA pool noise and $\phi$ is the strength of repression. As expected, $\tilde{\eta}_{\mu}$ plays a decisive role in determining the amount of extrinsic noise: the more variable the miRNA pool, the higher $\eta_{ext}$. Also, different miRNAs display different pool noise levels. $\tilde{\eta}_{\mu}$ estimates were similar among different constructs with the same miRNA binding sites, and they appeared to depend on miRNA repression strength. In fact, miRNA pool noise tends to decrease for highly repressive miRNAs. Moreover, the measured $\tilde{\eta}_{\mu}$ values were lower when the same miRNA was transcribed by multiple independent gene copies, suggesting that uncorrelated fluctuations in miRNA transcription average out.

In addition, protein expression noise appears to be reduced if miRNA-mediated regulation is combinatorial. Reporters with multiple miRNA binding sites displayed lower noise values than those regulated by a single miRNA. As mentioned, this reduction is further enhanced if the different miRNA pools are transcribed in an uncorrelated way. Since endogenous targets often contain several imperfect miRNA binding sites, the authors tested for this scenario by providing mCherry with multiple unrelated sites. This resulted in higher fold repression as compared to a non-combinatorially regulated case. Thus combinatorial regulation by miRNAs might reduce noise due to independent fluctuations compensating each other. Consistent with previous predictions, the overall noise was reduced, except when mCherry levels were high, and the $\eta_{int}$ dependence on $\sqrt{r}$ was confirmed. Therefore, although miRNA action displays opposing effects on intrinsic and extrinsic noise levels depending on the protein's expression level, the combination knocks down the overall noise at low expression and amplifies it at high expression with respect to unregulated protein production. However, according to \cite{farh2005widespread, sood2006cell}, miRNAs mostly target lowly expressed genes, that is they preferentially regulate those genes for which noise is more reduced upon miRNA-mediated regulation. Thus Schmiedel's findings suggest that the endogenous combinatorial regulation by miRNAs reduces $\eta_{tot}$ despite the additional extrinsic noise due to the variability of the miRNA pool.

Interestingly, Zare and co-workers \cite{zare2014evolutionarily} conducted a systematic analysis of the distribution of miRNA binding sites troughout the mouse genome, and they showed that such sites are found significantly more often within genes encoding fundamental regulatory proteins, especially those with high intrinsic transcriptional noise. These results appear to fit well into Schmiedel's idea of miRNAs reducing intrinsic noise while increasing extrinsic noise.
Ultimately, Schmiedel's work suggests an additional role for miRNAs in noise processing, potentially related to their presence in those biological processes that take advantage of gene expression variability, such as cell differentiation. 

MiRNAs have indeed been found to be largely involved in cell fate decision contexts, as reviewed in \cite{ivey2010micrornas}. Their role in differentiation has been investigated in the aforementioned work on PSCs \cite{Kumar2014} and in \cite{Klein2015}. Kumar et al. showed that miRNA knockdown results in significantly lower variability of the key PFs, thus driving the cell into a low-noise ground state, similarly to what observed when culturing cells in conditions that inhibit differentiation. However, miRNA knowdown cells seem to be more heterogenous than the latter and the authors suggests that this higher heterogeneity might be due to cells partly committing to the ground state. By profiling miRNA expression in PSCs, the authors highlighted the presence of two main miRNA groups, that is the ES-cell-specific cell-cycle regulating miRNAs (ESCC), well-known for being highly expressed in PSCs, and the let-7 miRNA family. By experimentally testing the exclusive and simultaneous expressions of the two miRNA groups and observing how gene expression was affected, they suggested that ESCC miRNAs can drive PSCs in a transition state where they are likely to differentiate, whereas let-7 alone appears to be able to repress a set of pluripotency genes, effectively leading to differentiation. Klein and co-workers showed that the intrinsic dimensionality of gene expression in pluripotent cells decreases after differentiation. A key role of miRNAs in mediating cell commitment to developmentally more advanced states by governing this fine-tuning is therefore suggested. Eventually, Garg and Sharp proposed that miRNAs may not only control cell-to-cell heterogeneity, but also generate it \cite{Garg2016}. Indeed, their suggestion is that miRNAs could enhance variability of the PFs through noise in the miRNA pool, in agreement with Schmiedel's idea. This noise could be transmitted on PFs through the regulatory network and PFs could in turn determine the miRNA expression profile, thus maintaining the cell in an established phenotypic condition. This idea would also be consistent with the several observations that miRNA profiles identify cell states.

Interestingly, the way miRNAs and their targets mostly interact is via titration, with the target responding in a threshold-linear fashion upon induction of its transcription rate or miRNA amount \cite{Levine2007, Mukherji2011, Bosia2013, Rzepiela2018}. The presence of a threshold-like behaviour defines mainly two regimes: a ``repressed'' (low target) regime, in which miRNA amount overcomes that of target, most of the targets are bound by miRNA molecules and the target is effectively repressed; and an ``unrepressed'' (high-target) regime, in which targets overcome miRNAs and there are enough free target molecules that can be translated \cite{Mukherji2011}, see Figure 1b. Around the threshold between the two regimes, where miRNAs and targets are highly coupled, the system is sensitive to fluctuations. This means that a fluctuation at the level of miRNA or target can propagate to other targets or miRNAs \cite{Bosia2013, Figliuzzi2013} (a phenomenon called {\it retroactivity} \cite{DelVecchio2008}), and closer the system is to the threshold, stronger is the retroactivity \cite{Bosia2017}. Since the steepness of the threshold between the two regimes depends on the interaction strength between miRNA and target, if the steepness is high (i.e. strong interaction), what may happen is that small intrinsic fluctuations induce single cells to sample the two regimes, thus giving bimodal distributions on the target at the population level. This said, the presence of extrinsic noise - as that in the miRNA pool - facilitates this sampling. Indeed, the broader the noise, i.e. the broader the miRNA distribution, the easier to have values of miRNA such that the target is for one cell in the repressed regime and for another cell in the unrepressed one. Bimodal distributions may then appear even if the interaction strength is mild \cite{DelGiudice2018role, DelGiudice2018stochastic}. Such a scenario is valid not only for a one-miRNA/one-target system but also when multiple miRNAs and targets are interacting. In this situation, indeed, it is still possible to define a threshold around which all the targets and miRNAs are coupled, with the strength of these couplings determined by the particular interaction strengths \cite{Bosia2013}, see Figure 1c. The net effect is that several targets can simultaneously display bimodal distributions, thereby allowing the emergence of multiple phenotypic configurations, each defined by a combination of target states.

It is known that cell types are identified by a small set of miRNAs that dominates the total miRNA pool (``master miRNAs'') \cite{landgraf2007mammalian, marson2008connecting}. Suzuki et al. \cite{suzuki2017super} analyzed the relationships between master miRNAs and regulatory regions called ``super-enhancers'' (SEs), known for controlling cell identity. SEs were found to be connected to a few highly abundant miRNAs, which turned out to be the previously identified master miRNAs. Moreover, SEs were observed to widely shape miRNA expression. Since the interplay between master miRNAs and SEs appears to indentify cell state, it is suggested that these miRNAs play a fundamental role in transitions between cell states, i.e. differentiation processes, and SEs might act as noise-generators by enhancing the miRNA pool noise, thereby favouring the emergence of bimodal phenotypes.
These results suggest that miRNA could increase the cell-to-cell variability of their targets. If the targets are key developmental factors such as morphogens, this variability can be the trigger of cell state transitions \cite{chakraborty2019}.
These experimental works along with theoretical modelling showed that quantitative investigation is crucial for understanding the impact of miRNAs in managing noise. Yet, current studies are often limited by difficulties in exactly quantifying the molecular interactions between miRNAs and their targets. These limitations and the recent advances in this field are discussed in the next section.

\section{Quantitative inference on miRNA-target interactions}

In order to understand how miRNAs deal with gene expression noise, the combination of experiments and theoretical modelling of miRNA-target interactions provides an essential tool. Consistent parameter estimates allow precise quantitative predictions on expression variability. Yet, theoretical modelling of biological interactions is often built on network theory. Therein, the gene-expression machinery is described as a set of nodes representing molecules such as miRNAs, mRNAs and proteins which are connected by links representing the interactions amongst them, see Figure 1d. It stems clearly out that this kind of approach may normally require a high number of coarse-grained parameters to be defined. In fact, a long lasting question related to the parameters modulating miRNA-mediated processes is to what extent their values influence the processes' outcomes, a question that falls on parameter estimate. Precise quantitative estimates would indeed improve both the algorithms aimed at predicting the target genes of specific miRNAs as well as the understanding of the biological mechanisms underlying experimental observations, therefore increasing the predictive power of theoretical models.

In the last few years, mathematical modelling of miRNA-target interactions has greatly focused on the aforementioned interaction networks involving miRNAs, TFs and target genes, as reviewed in  \cite{Cora2017, ferro2019endogenous}. Laurenti and colleagues focused on theoretical interaction circuits aimed at reducing noise, that is circuits that act as molecular filters \cite{laurenti2018molecular}, by using a network-theory approach. These networks are constituted by simple combined biological interactions, such as the co-expression of two species that subsequently bind together, i.e. the iFFL. The authors' suggestion, which follows what pointed out by Riba and coworkers few years before \cite{riba2014}, is that these molecular filters are pervasive in gene expression and that miRNAs participate in such modules. In fact, a few miRNA-mediated noise-reducing networks have been proven to be overrepresented in mammalian genomes by Tsang et al. \cite{tsang2007microrna}.

An important example in the framework of miRNA-mediated network motifs has been brought forward by Lai and colleagues \cite{lai2016understanding}. Therein, the role of miRNA-mediated regulatory circuits in fine-tuning gene expression by buffering noise was elucidated by theoretical means such as Ordinary Differential Equation (ODE) modelling. For instance, the endogenous feedback loop formed by E2F1 and the miR-17-92 family was shown to display bistability, that is the E2F1/miR-17-92 can only switch between ``ON/OFF'' and ``OFF/ON'' states. Only a crucial drop in the upstream E2F1-inducing signal can cause the system to switch to the opposite state, thus this type of network is inherently robust to fluctuations \cite{osella2014interplay}. However, this kind of circuits are also known to be importantly involved in transitions between cell types. For instance, the same loop has been found to regulate the epithelial to mesenchymal transition, where the two phenotypes correspond to the two bistable states mentioned above. Indeed, the miRNA level determines in which state the system will collapse, thus an increased noise in miRNA expression could act as a trigger for the transition. In fact, when involved in feedback loops, miRNAs have been shown to increase variability if required for achieving cell differentiation or cell state changes \cite{Herranz2010}.

The iFFL's noise-buffering properties have also been widely studied in a number of theoretical and experimental works \cite{bosia2012, osella2011, Strovas2014, Grigolon2016}. Results have highlighted its ability to adapt to transient signal changes, that is the target gene's expression level displays little susceptibility to upstream fluctuations for a wide parameter range.
Nevertheless, many quantitative aspects of these interactions, i.e. how noise is affected by the parameter regime, are still not fully understood. Carignano and co-workers tried to find an answer to this issue by searching for the iFFL's parameter regimes where noise is most efficiently buffered \cite{carignano2018extrinsic}. They demonstrated that if extrinsic noise is static, miRNA-mediated translational inhibition rejects noise for a broader parameter range than protein decay amplification. As of dynamic extrinsic noise, a special case of the iFFL where the target gene and the miRNA are transcribed together was shown to either reduce or amplify product variability depending on the relationship between the timescale of extrinsic fluctuations and that of mRNA and miRNA degradation.

In general, miRNAs' characteristic timescales of biogenesis, action and decay are of course crucial in determining a network's outcome. Since miRNA expression does not require protein synthesis, miRNAs were generally viewed as fast regulators of gene expression compared to transcription factors \cite{shimoni2007regulation, hobert2008gene}. However, by combining theoretical modelling with miRNA induction and transfection datasets, Hausser and co-workers showed that the timescale of miRNA-mediated regulation is slower than expected \cite{Hausser2013}. Indeed, miRNAs only function as part of complexes with Argonaute (Ago) proteins \cite{Dueck2012}, with the concentration of miRNA-Ago complexes usually considered constant \cite{Khan2009}. Thus, the commonly observed small changes in protein levels seem to be due to both delays in miRNA loading into Ago proteins and to the slow protein decay.

Huge effort was also spent in quantifying the strength of miRNA-target interactions, which represents a pivotal quantity for miRNA-target prediction algorithms \cite{Khorshid2013} and an important parameter in theoretical models \cite{Bosia2013, Figliuzzi2013}. Having in mind the miRNA-induced linear-threshold target behaviour mentioned above, the affinity of a miRNA and its target determines the steepness of the threshold, and thus the susceptibility of the target to fluctuations in the amount of miRNAs or in the amount of other endogenous targets competing for the same miRNAs. Wu and colleagues \cite{wu2018mirna} worked on mutations in the miRNA-binding mRNA sequences: they quantified each binding energy change of 67159 different mutations. Dealing with 21 cancer types, they showed that the higher the loss of strength, the more expressed were the cancer-related genes. As mentioned, miRNA-target affinity determines the extent to which miRNA affects mRNA translation compared to its degradation. Thus these results suggest that poor mRNA degradation may be a determinant factor in cancer. 

With a series of seminal papers, Zavolan's and van Nimwegen's groups moved as well in the direction of uncovering miRNA-target strengths of interaction \cite{Khorshid2013, Breda2015, Rzepiela2018}. It is worth discussing this work a little deeper in order to exemplify how a quantitative study on miRNA-target interactions involving theoretical and experimental tools can be performed. The authors first defined a model-based method to infer perfectly and unperfectly complementary miRNA targets, i.e. canonical and non-canonical sites, from Argonaute 2 cross-linking and immunoprecipitation data \cite{Khorshid2013}. The model (MIRZA) includes parameters related to base pairs, loops in the sequences and position-dependent energy constraints imposed by Argonaute proteins. With these parameters, MIRZA computes the energy of a miRNA-mRNA hybrid, which allows calculating the frequencies of RISCs binding to each miRNA in a pool of different miRNAs. Parameters are then inferred from Ago-CLIP data collected in HEK293 cells by maximising the binding probabilities of mRNA fragments observed in the samples. 

The inference procedure is performed by calculating a ``target quality'' $R(m|\mu)$ that gives the affinity of each miRNA $\mu$ with each mRNA fragment $m$, which can also be read as the fraction of frangment $m$ among target sites bound to miRNA $\mu$. This quantity is obtained by summing over all possible hybrid structures that $m$ can form when binding $\mu$. The fraction of time that $m$ is bound to a RISC loaded with miRNA $\mu$ is proportional to $R(m|\mu)\pi_{\mu}$, where $\pi_{\mu}$ is the total fraction of $\mu$-loaded RISC bound to mRNA. These fractions, called miRNA priors, are inferred from each CLIP dataset. The total probability of fragment $m$ being bound to miRNA is $R(m)=\sum_{\mu}{R(m|\mu)}\pi_{\mu}$ and the total likelihood of a dataset is $R(D)=\prod_{i}{R(m_i)}$. Results on parameters capture several already known features of the miRNA-mRNA bound. For instance positions 2-7 of binding sites, commonly known as the seed region, have the largest contribution to the energy, and multiple other predictions on nucleotides depending on their position come out. 

By applying the model with fitted parameters, it is possible to predict which miRNA $\mu$ is more likely to bind each fragment $m$, and even the structure of the most likely miRNA-mRNA hybrid. What has emerged from such predictions is that non-canonical sites are bound to a larger extent to miRNAs more bound to RISCs, that is miRNAs with higher $\pi_{\mu}$. In other words, $\pi_{\mu}$ correlates positively with $\mu$'s expression. Thus lowly expressed miRNAs target sites with high affinity, whereas highly expressed miRNAs also target low-affinity sites. To test the effectiveness of predicted sites, mRNA fold changes estimated by MIRZA were validated by comparing them to those measured upon miRNA transfection. MIRZA predicts the existence of many functional non-canonical sites that had not been previously found by other miRNA-target interaction models. Moreover, they are found to be evolutionarily conserved, as their presence is significantly larger than expected by chance. The authors suggest that MIRZA could be further improved by adding conservation information to the model. 

In a subsequent work, MIRZA was used to quantify the strength of miRNA-target interactions \cite{Breda2015} and the results showed that the computationally predicted binding energies strongly correlate with the energies estimated from biochemical measurements of Michaelis-Menten constants. Single-cell RNA-seq analysis then opened the way to infer parameters describing the response of even hundreds of miRNA targets and ideally verify predictions that were only possible in a theoretical framework. A great example of such quantitative estimation is the work of Rzepiela and co-workers \cite{Rzepiela2018}. There, the authors inferred the sensitivity of individual targets to miRNA regulation from their expression in cells with varying miRNA level. Results showed that the response of miRNA targets to miRNA induction is hierarchical: the targets of a miRNA can be ordered in a hierarchy based on the miRNA concentrations at which they respond within the endogenous context of all other miRNAs and targets in the cell. Specifically, the few targets with higher Michaelis-Menten constants displayed higher sensitivity to changes in miRNA amount. Moreover, responses followed behaviours that were theoretically predicted in \cite{Bosia2013, Figliuzzi2013}. 

Quantifying target response to miRNA-mediated regulation can make a decisive contribution in the comprehension of miRNA-mediated noise. Indeed, as shown by Schmiedel and colleagues \cite{Schmiedel2015}, intrinsic noise is related to fold repression. Thus quantitative estimations on target sensitivity to miRNA induction such as the ones by Rzepiela et al. \cite{Rzepiela2018} can be of great value for the understanding of such variability. Also the inference of miRNA-target strengths of interaction can be used to improve predictions on miRNA-mediated gene expression noise, as shown in a recent work \cite{Bosia2017}. It is indeed observed that in the sole presence of intrinsic noise, with combinatorial miRNA-target interactions, an important parameter governing cell-to-cell variability appears to be the interaction strength, with the latter proportional to the number of miRNA binding sites. For instance, target bimodality is achieved only for high strength of interaction values. A subsequent theoretical study by Del Giudice et al. \cite{DelGiudice2018role} investigated the relationship between extrinsic noise, target response bimodality and miRNA-target affinity. The results suggested that if extrinsic noise is added to the system, target bimodality appears also when the miRNA-target interaction strength is small, with the size of the bimodality range again dependent on this parameter. 

Another interesting quantitative aspect is the extent to which miRNAs affect mRNA decay compared to their translation rate. The general idea was that miRNAs affect the messenger decay rate more than they affect translation \cite{Eichhorn2014}. According to this, one would expect changes in mRNA levels and in protein levels to be strongly correlated. Instead, mRNA and protein amount variations seem uncoupled, with the repression of target translation preceding the increase in its degradation rate and the protein amount typically changing less than that of the mRNA \cite{Bazzini2012, Hausser2013, Eichhorn2014}. However, theory suggests \cite{gedeon2012delayed} that the observed uncorrelation might be partially explained by a delay due to miRNA maturation \cite{Hausser2013}.
Bearing in mind that the way miRNAs preferentially affect protein production depends on miRNA-target affinity, and is thus related to interaction strength, the estimation of parameters that measure the impact of miRNAs on target degradation against translation could also play a role in the predictions on miRNA-mediated noise.

\begin{figure}[!h]
\begin{center}
\includegraphics[scale=0.5]{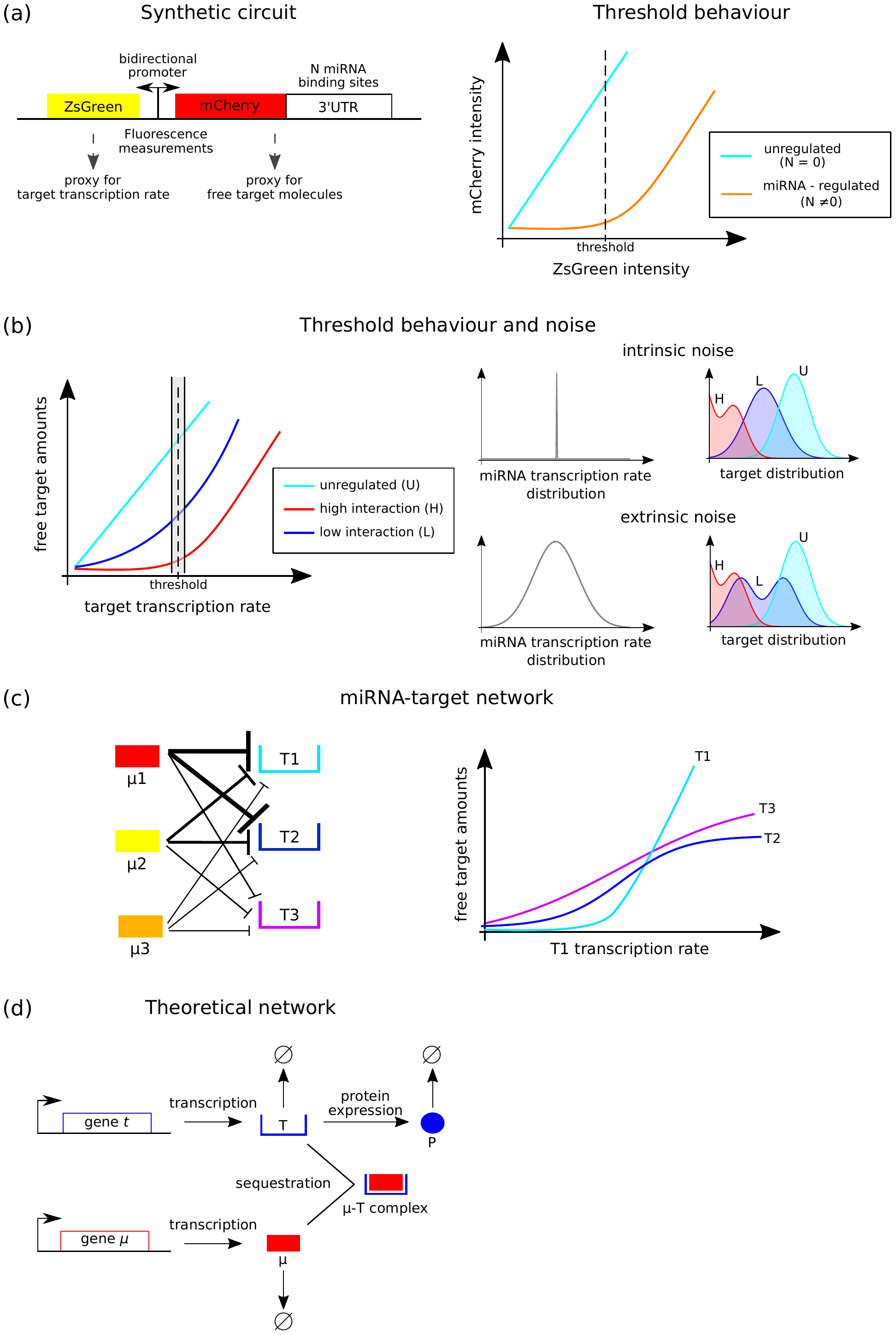}
\end{center}
\caption{(a) Synthetic circuit built by Schmiedel et al. \cite{Schmiedel2015}. The circuit consists of a bidirectional plasmid encoding two copies of a gene, one of which contains a number N of miRNA binding sites in its 3'UTR (ZsGreen and mCherry). Gene transcript levels are quantified through fluorescence measurements. The unregulated gene transcript level ZsGreen can be considered as a proxy for transcriptional activity. The miRNA-regulated transcript level, that is the mCherry intensity, can thus be compared to the unregulated one in order to quantify the effects of miRNA-mediated repression. The target gene transcript is ``sequestered'' by the miRNA as described in (d), thus the level of miRNA-regulated mRNA (mCherry) can be assumed as a proxy for the amount of free target transcript. The qualitative plot on the right represents the amount of free mRNA mCherry as a function of ZsGreen amount. The cyan line represents the case where both ZsGreen and mCherry are devoid of miRNA binding sites in their 3'UTR (N $=$ 0), while the orange line describes the miRNA-regulated mCherry case (N $\neq$ 0). As expected, the first scenario results in a linear relationship between ZsGreen and mCherry amounts. By contrast, in the second scenario the mCherry sequestration by the miRNA generates a threshold behaviour: below the threshold the free mCherry level is lower than the amount of ZsGreen, whereas above the threshold it overcomes the ZsGreen level. Adapted from \cite{Schmiedel2015}. 
(b) Schematic representation of the bimodal distributions obtained when combaining threshold-like response and noise. The grey shadowed region around the threshold identifies a transcription rate for which the target may be bimodal in case of pure intrinsic noise (upper left panel) or extrinsic noise in the miRNA pool (lower left panel). With intrinsic noise only, a high miRNA-target interaction strength is necessary to have bimodal target (red line), while with extrinsic noise bimodality is present also for mild interactions (blue line).
(c) Schematic example of a miRNA-target regulatory network with the associated threshold-like behaviour. All miRNAs act as repressors of all target mRNAs but with different strengths of interaction, which are represented by the different thicknesses of the links. Adapted from \cite{Bosia2013}. (d) Theoretical circuit representing the transcription and interaction of a miRNA and its target. The target is transcribed from gene \textit{t} into mRNA transcript T. T can be degraded, translated into the protein P (which can be degraded as well) or sequestered by the miRNA. The miRNA is transcribed from gene $\mu$. The corresponding miRNA transcript $\mu$ can either be degraded or form a complex with its target mRNA T.}
\end{figure}

\section{Conclusions}

In this review, we discussed the literature underlying the recent efforts in the understanding of miRNAs' role in target noise control. While in the past miRNAs were believed to mainly act as noise bufferers, more recent works suggest that extrinsic sources of noise, such as fluctuations within miRNA pools, lead to target noise increase, possibly driving the formation of different phenotypes. Thus, altogether, the recent works shed light on the possibility that living systems do not function by only minimising stochasticity but that they have instead evolved by optimising the possible effects of randomness. Also, they highlight the importance of interdisciplinary approaches in defining directions for the quantitative identification of the optimisation mechanisms orchestrating life. In this respect, the parallel advancing of inference methods to quantitatively estimate parameters related to miRNA-target interactions from theoretical modelling is of extreme importance. Indeed, parameter estimation is the only way to precisely predict expression variability. In this direction, the studies revised in the last section may all be used to improve the predictions on miRNA-mediated noise and hopefully pave the way for model-based therapeutic perspectives, with a constant interdisciplinary approach, and more in general for the understanding of the hidden secrets of living systems.

\section{Acknowledgments}

This work was supported by The Francis Crick Institute which receives its core funding from Cancer Research UK (FC001317), the UK Medical Research Council
(FC001317), and the Wellcome Trust (FC001317), to S.G.

\bibliographystyle{ieeetr}

\end{document}